\documentclass[a4paper,12pt]{article}
\usepackage{amssymb}
\usepackage{latexsym}
\usepackage[dvips]{graphicx}
\usepackage{cite}
\newcommand{\be}{\begin{equation}}
\newcommand{\ee}{\end{equation}}
\newcommand{\bea}{\begin{eqnarray}}
\newcommand{\nn}{\nonumber}
\newcommand{\eea}{\end{eqnarray}}

\begin{document}

\begin{titlepage}
\mbox{} \vspace{.2in}
\begin{flushright}
UB-ECM-PF-03/16
\end{flushright}
\begin{centering}
\vspace{.7in}
{\Large{\bf 2d Stringy Black Holes and Varying Constants}}
\\

\vspace{.5in} {\bf  Elias C.
Vagenas\,\footnote{evagenas@ecm.ub.es} }\\

\vspace{0.3in}

Departament d'Estructura i Constituents de la Mat\`{e}ria\\
and\\ CER for Astrophysics, Particle Physics and Cosmology\\
Universitat de Barcelona\\
Av. Diagonal 647\\ E-08028 Barcelona\\
Spain\\
\end{centering}

\vspace{0.7in}
\begin{abstract}
Motivated by the recent interest on models with varying constants
and whether black hole physics can constrain such theories,
two-dimensional charged stringy black holes are considered. We
exploit the role of two-dimensional stringy black holes as toy
models for exploring paradoxes which may lead to constrains on a
theory. A two-dimensional charged stringy black hole is
investigated in two different settings. Firstly, the
two-dimensional black hole is treated as an isolated object and
secondly, it is contained in a thermal environment. In both cases,
it is shown that the temperature and the entropy of the
two-dimensional charged stringy black hole are decreased when its
electric charge is increased in time. By piecing together our
results and previous ones, we conclude that in the context of
black hole thermodynamics one cannot derive any model independent
constraints for the varying constants. Therefore, it seems that
there aren't any varying constant theories that are out of favor
with black hole thermodynamics.
\end{abstract}
\end{titlepage}
\newpage

\baselineskip=18pt
Despite the fact that the idea of varying constants is more than
forty years old
\cite{dirac,jordan1,jordan2,teller,dicke1,dicke2,brans} there has
not been an experimental evidence as yet. The theoretical
investigations that have been performed on this idea, can be
roughly classified in two groups : (a) works in the framework of
four-dimensional spacetimes where the fundamental constants were
varying in space and/or in time, (b) and those works in higher
dimensional spacetimes where the four-dimensional effective
constants  depend on any temporal or spatial variation of the
structure or the size of extra-dimensions\,\footnote{The work
presented here can be viewed within this framework. In particular,
we treat two-dimensional black holes which are derived from a
string theory heterotically compactified to two dimensions.
Therefore, one can expect the time dependence of the
two-dimensional constants to be a result of this heterotic
compactification. Generally speaking, one cannot simply ``write
in'' variations of constants since it is possible  when a constant
varies the black hole solution no longer exists \cite{joao}. }.
Regarding the former case, representative works are those of
Bekenstein who treated the variability of the fine structure
constant through a spacetime varying electron charge \cite{bek1},
and Moffat \cite{moffat1,moffat2} who in order to solve  some
cosmological problems introduced a time varying speed of light.
Regarding now to the latter case and specifically within the
context of string theory, the presence of the massless dilaton
field which determines the string coupling constant
$g_{s}=e^{\phi/2}$ and thus is the link between gravitation and
matter interactions, led to violation of the universality of free
fall and a time variation of the fine structure
constant\,\footnote{Recently, Bekenstein developed a mechanism
which prevents equivalence principle violations due to variations
of the fine structure constant $\alpha$, to be measurable
\cite{bek2}. This compensating mechanism was introduced in the
general framework discussed in \cite{bek1}. }. Damour and Polyakov
in an attempt to reconcile massless dilaton field with the
experimental tests of the equivalence principle presented a
decoupling mechanism \cite{damour1,damour2}.
\par
 Recent astronomical observations revived again the possibility of varying
constants. Actually, there were hints that the fine structure
constant $\alpha= e^{2}/\hbar c$ is increasing in time
\cite{webb1,webb2,webb3}. Precisely, by developing a new more
sensitive (compared to the older ``alkali-doublet'') method called
``many-multiplet'', there has been a statistical evidence for a
smaller $\alpha$ with $\Delta \alpha /\alpha=\left(- 0.72\pm
0.18\right) \times 10^{-5}$ for $z\approx 0.5 - 3.5$.
Additionally, there were hints for a time variation of the
electron to proton mass ratio $\mu=m_{e}/m_{p}$
\cite{ivan1,ivan2}. In particular, by measuring the $H_2$
wavelengths in the high-resolution of two quasars with damped
Lyman-$\alpha$ systems at $z=2.3377$ and $z=3.0249$, they detected
a time variation of $\mu$ with $\Delta \mu /\mu=\left(-5.7\pm
3.18\right) \times 10^{-5}$ to be the most conservative result.
The exact expression relating the fine structure constant $\alpha$
with the electron to proton mass ratio $\mu$ is still lacking but
within the above-mentioned context presented by Damour and
Polyakov the masses of electron and proton, and  the fine
structure constant depend on the massless dilaton field and hence
are related. It should be noted that both measurements are of
equivalent importance since they are non-zero detections.
\par\noindent
 The aforesaid astronomical results gave an impulse to new theoretical
 works on varying constants\,\footnote{An excellent review
concerning the fundamental constants and their variability is
provided by Uzan \cite{uzan}.}.
 On one hand, expanding the previous works of Bekenstein and Moffat, Albrecht and Magueijo \cite{alb}
 presented as an alternative to Standard Big Bang model of the
Universe in order to solve cosmological puzzles, a model of
temporal varying speed of light (see also
\cite{barrow1,barrow2,moffat3,barrow3,barrow4,barrow5}). On the
other hand, Damour, Piazza and Venezianno obtained to extend the
model of Damour and Polyakov \cite{damour1,damour2} in a way that
the coupling functions which depend on the massless dilaton field,
have a smooth finite limit for the infinitely large values of the
bare string coupling \cite{damour3,damour4}.
\par\noindent
In an attempt to test theories of varying-$c$ or varying-$e$,
Davies, Davis and Lineweaver \cite{davies} considered as testing
ground the black hole thermodynamics.  They argued that the
entropy of a four-dimensional Reissner-Nordstr\"{o}m black hole
solution of Einstein's theory of gravity decreases if the electric
charge $Q$ of this black hole increases while the Newton's
constant $G$ and $c$ are kept constant. On the contrary, an
increase of $c$ will not lead to a violation of the second law of
black hole thermodynamics\,\footnote{The compatibility of varying
$c$ cosmologies with the (generalized) second law of
thermodynamics was investigated in \cite{luis,youm}.}. Therefore,
they concluded that theories in which the electric charge $e$
varies in time, are disfavored since they violate the second law
of thermodynamics. Immediately afterwards, Carlip and Vaidya
\cite{carlip1} showed that when one considers the full thermal
environment of the four-dimensional Reissner-Nordstr\"{o}m black
hole then no such conclusion is extracted. Fairbairn and Tytgat
\cite{fair} also proved for the four-dimensional electrically
charged black hole solution of string theory (well-known as GHS
black hole) that its entropy remains constant with respect to
adiabatic, i.e. slow, variations of the fine structure constant
$\alpha$, irrespectively of whether the change is due to an
increase of $e$ or an decrease of $c$.
\par\noindent
In this paper, we consider the two-dimensional charged stringy
black hole of McGuigan, Nappi and Yost \cite{nappi1}. Our
motivation for this choice is the fact that the two-dimensional
stringy black holes provide the simplified framework in which one
can explore paradoxes such as the Hawking effect, that may lead to
constrains on the theory \cite{harvey}. Our starting point will be
the two-dimensional effective action realized in heterotic string
theory \cite{nappi2}\be S= \int d^{2}x
\sqrt{-g}e^{-2\phi}\left[R+4(\nabla\phi)^{2}+4\lambda^{2}-
\frac{1}{4}F^{2}+\nabla^{a}\left(4e^{-2\phi}\nabla_{a}\phi\right)
\right]\label{action1}\ee where $g$ is the determinant of the
two-dimensional metric $g_{\mu\nu}(x)$, $\phi$ is the dilaton,
$\lambda^{2}$ is the cosmological constant and $F_{\mu\nu}$ is the
Maxwell stress tensor.
\par\noindent The line element of the two-dimensional charged
stringy black hole solution derived by McGuigan, Nappi and Yost
from action (\ref{action1}) is given, in the ``Schwarzschild"
gauge \cite{lee,elias}, as \be ds^2 = -g(r)dt^2 + g^{-1}(r)dr^2
\label{linelement2} \ee where \be g(r) = 1 - 2 m e^{-2\lambda r} +
q^{2} e^{-4\lambda r} \label{rnmetric} \ee and the dilaton field
is given as \footnote{Apart from viewing equation (\ref{dilaton1})
as a solution of action (\ref{action1}) for the dilaton field, one
can also make two more comments for the linear dependence of
dilaton field on the spatial coordinate. Firstly, the ``linear''
dilaton field is a common feature for lower-dimensional string
theories \cite{kummer}. For instance, in the case of CGHS black
holes for zero mass ($M=0$), one gets the ``linear'' dilaton
vacuum which in the presence of any matter is perturbed and hence
a formation of a black hole takes place \cite{callan,russo}.
Secondly, one can treat the dilaton field as the ``radial''
coordinate of the two-dimensional spacetime since it is linearly
related to the spatial coordinate. For instance, the
two-dimensional entropy can be written in terms of the value of
the dilaton on the black hole horizon instead of the value of the
radial coordinate \cite{cadoni}. } \be \phi(r)=\phi_{0}-\lambda
r\label{dilaton1}\ee with $0<t<+\infty$, $r_+<r<+\infty$, $r_+$
being the future event horizon of the black hole.
\par\noindent
The constants $m$ and $q$ are related to the (ADM) mass $M$ and
electric charge $Q_{el}$ of the two-dimensional charged stringy
black hole, respectively, as \be M= 4\lambda m
e^{-2\phi_{0}}\label{mass}\ee \be Q_{el}=2\sqrt{2}\lambda q
e^{-2\phi_{0}}\label{charge}\ee where these values have been
evaluated at infinity, i.e $r\rightarrow \infty$. \par\noindent It
can be easily seen from action (\ref{action1}) that the fine
structure constant is given as \be \alpha=
e^{2\phi_{0}}\label{alpha1}\ee where $\phi_{0}$ is the asymptotic
value of the dilaton field. \par\noindent At this point it should
be pointed out that the fine structure constant as defined in four
dimensions, i.e. \be \alpha=\frac{e^{2}}{\hbar
c}\label{fsc}\hspace{1ex},\ee is no more dimensionless in the two
dimensional spacetimes\,\footnote{This is probably due to the
super-renormalizability of QED in two dimensions \cite{ghosh}.}.
In particular, for the two-dimensional black hole backgrounds
under consideration, since the metric function, the dilaton field
and the action are considered  as dimensionless one gets \bea
[\lambda]&=&L^{-1}\,T^0\nn\hspace{1ex},\\
  {[} M]&=& L^{-1}\,T^0\nn\hspace{1ex},\\
{[}Q_{el}]&=&L^{-1}\,T^0\nn \hspace{1ex},\eea where $[\,]$ denotes
the dimension and $L$, $T$ denote the length and time,
respectively. It is noteworthy that in two dimensions the Newton's
constant $G_{N}$ is dimensionless and thus it  is not possible to
define even the  ``Planck'' length. \par\noindent The velocity of
speed of light has dimensions  \be [c]=L\,T^{-1}\nn\ee while the
Planck's constant has dimensions \be
[\hbar]=L\,T^{-1}\hspace{1ex}.\nn\ee Thus, we define the
two-dimensional fine structure constant\footnote{At this point, it
should be stressed that although the two-dimensional fine
structure constant introduced here is a link between gravitation
and the electromagnetic properties of the two-dimensional black
hole, it is not a priori related to atomic physics as the
four-dimensional one. Thus, there is no obvious reason to relate
the two-dimensional fine structure constant with the one we
measure in the astrophysical systems.} in analogy to four
dimensions, by using the length scale $1/\lambda$, as follows\be
\alpha=\frac{ \hbar \,e^2}{\lambda^{2}\, c}\hspace{1ex}.\ee
\par\noindent Pursuing a parametrization analogous to the
four-dimensional case  the metric function (\ref{rnmetric})
factorizes as \be g(r)=(1-\rho_- e^{-2\lambda
r})(1-\rho_+e^{-2\lambda r}) \ee where \be \rho_\pm= m  \pm
\sqrt{m^2-q^2} \label{root2}\hspace{1ex}. \ee It is easily seen
that the outer event horizon $H^+$ is placed at the point
$r_+=\frac{1}{2\lambda}ln\rho_+$, while the ``inner" horizon $H^-$
is at the point $r_-=\frac{1}{2\lambda}ln\rho_-$.
\par\noindent The temperature of the two-dimensional charged stringy black
hole can easily be derived by implementing its definition
\cite{birrell} \be
T_H=\frac{\kappa}{2\pi}\hspace{.2in}\mbox{and}\hspace{.2in}
{\kappa}=\frac{1}{2}\left. \frac{\partial g(r)}{\partial
r}\right|_{r=r_+} \ee which yields the following expression for
the Hawking temperature \cite{elias} \be
T_H=\frac{\lambda\sqrt{m^2 -q^2}}{\pi\left(m+\sqrt{m^2
-q^2}\right)} \label{temp}\hspace{1ex}. \ee
\par\noindent
From thermodynamics \cite{nappi2}, we can also obtain the entropy
of the two-dimensional charged stringy black hole \bea S&=&4\pi
e^{-2\phi_{0}}\left(m+\sqrt{m^2 -q^2}\right)\nn\\
&=&4\pi m e^{-2\phi_{0}}\left(1+\sqrt{1 -
\gamma}\right)\label{eg}\eea where
\be\gamma=\left(\frac{q}{m}\right)^{2}\label{gamma}\ee is a
dimensionless parameter which in terms of the mass $M$ and the
electric charge $Q_{el}$ of the two-dimensional charged stringy
black hole is given as \be
\gamma=2\left(\frac{Q_{el}}{M}\right)^{2}\ee and the electric
charge $Q_{el}$ is quantized in units of the electric charge $e$,
i.e. $Q_{el}=n e$ \footnote{Up to now, a full theory of quantum
gravity does not exist thus black hole mass is not treated as
quantized, i.e. $M=\sqrt{r} M_{pl}$ ($r$ is discrete here), and
furthermore the mass quantization may lead to a discrete evolution
of the fine structure constant $\alpha$ \cite{carlip2}.}. It is
obvious that the temperature (\ref{temp}) of the two-dimensional
charged stringy black hole can now be written in terms of the
dimensionless parameter $\gamma$ as \be
T_{H}=\frac{\lambda}{\pi}\left(1+\frac{1}{\sqrt{1-\gamma}}
\right)^{-1}\label{tg}\hspace{1ex}.\ee It should be pointed out
that the entropy (\ref{eg}) and the temperature (\ref{tg}) have
both been evaluated on the outer horizon, i.e. $r=r_{+}$, of the
two-dimensional charged stringy black hole. Since we are going to
consider the increase of the fine structure $\alpha$ solely due to
an increase in the electric charge $e$ all other quantities are
kept constant. It is worthy to note that the reason for
introducing the dimensionless parameter $\gamma$ in the
expressions for the temperature and the entropy of the
two-dimensional charged stringy black hole is the fact that one
can only consider variation of dimensionless
constants\,\footnote{Lately, there was a debate on whether is
physically meaningful to consider time variation of dimensional
constants  \cite{duff1,moffat4}.}. Thus, by varying $e$ we mean
here the variation of dimensionless quantities that depend on $e$.
\par\noindent
It is obvious from expression (\ref{gamma}) that an increase of
the electric charge $e$ implies an increase in $\gamma$. Thus,
concerning the temperature of the two-dimensional stringy black
hole, it is easily derived from expression (\ref{tg}) that an
increase in $\gamma$ causes the temperature to decrease.
Therefore, as the electric charge $e$ increases the
two-dimensional stringy black holes becomes cooler. It can also be
checked from expression (\ref{eg}) that an increase in $\gamma$
will lead the entropy of the two-dimensional charged stringy black
hole to become smaller. Therefore, an increase in the electric
charge $e$ in the specific black hole background is at the risk of
violating the second law of black hole thermodynamics.
At first sight, this seems to be catastrophic. Consequently, one
is considering to rule out varying $e$ theories in order to avoid
the aforesaid violation. However, since we are concerned with
physical observations, one should not be interested in studying an
isolated black hole rather than a black hole in its thermal
environment. Therefore, we now consider the two-dimensional
charged stringy black hole contained in a ``box'' and the physical
measurements will be made on the boundary of this black hole
spacetime, i.e. on the ``wall'' of the ``box'' which is located at
a finite distance of the radial coordinate. It is convenient to
follow the terminology of Gibbons and Perry \cite{perry} and the
action is now given as \be S= \int_{\mathcal{M}} d^{2}x
\sqrt{-g}e^{\phi}\left[R+(\nabla\phi)^{2}+4\lambda^{2}-
\frac{1}{4}F^{2}\right]+2\int_{\partial \mathcal{M}}e^{\phi }K
d\Sigma\label{action2} \ee where $K$ is the trace of the second
fundamental form of the boundary, i.e. $\partial \mathcal{M}$, of
the two-dimensional spacetime $\mathcal{M}$ and $d\Sigma$ is the
volume on the boundary. It is evident that the fine structure
$\alpha$ takes the form \be \alpha=e^{-\phi_{w}}\label{alpha2}\ee
where $\phi_{w}$ is the value of the dilaton field on the
boundary, i.e. the ``wall'' of the ``box''. The line element is
given, in the unitary gauge \cite{elias}, by \be
ds^2=-\frac{\left(m^2 -q^2 \right) \sinh^2(2 \lambda
y)}{\left(m+\sqrt{m^2 -q^2} \cosh^2(2 \lambda y)\right)^2}dt^2 +
dy^2 \ee where the unitary variable is given by \be
y=\frac{1}{\lambda}\ln\left[
\sqrt{\frac{1}{\mu}(e^{2\lambda(r-r_+)}-1)} +
\sqrt{\frac{1}{\mu}(e^{2\lambda(r-r_+)}-1)+1} \right] \ee and
$0<y<+\infty$, while $\mu = 1-\frac{\rho_-}{\rho_+}$.
\par\noindent The dilaton field evaluated on the wall, i.e. $y=y_{w}$, will be given as \be
\phi_{w}=\phi_{0}+\ln\left[\frac{1}{2}\left(\frac{1}{\sqrt{1
-\gamma}}+\cosh(2 \lambda y_{w} )\right)\right]\label{dilaton2}\ee
while the dilaton charge is \be
D=\frac{1}{2}e^{\phi_{0}}\left(\frac{1}{\sqrt{1-\gamma}}+\cosh\left(
2\lambda y_{w}\right)\right)\label{dcharge} \hspace{1ex}.\ee The
electric charge $Q_{el}$ and the (ADM) mass $M$ of the
two-dimensional charged stringy black hole evaluated inside the
``box'' are given, respectively, as \bea M&=&\frac{2\lambda
}{\sqrt{1-
\gamma}}e^{\phi_{0}}\\
Q_{el}&=&\frac{\sqrt{2}\lambda }{m}\frac{\sinh \left(2\lambda
y_{w}\right)}{\left[1+\left(1-\gamma\right)\cosh\left(2\lambda
y_{w} \right)\right]}e^{\phi_{0}}\hspace{1ex}.\eea The local
temperature of a self-gravitating system in thermal equilibrium is
given by Tolman's law as \be
T_{local}=\frac{T_{H}}{\sqrt{-g_{tt}}}\hspace{1ex},\ee therefore
the temperature of the two-dimensional charged stringy black hole
evaluated on the ``wall'' of the ``box'' is \be
T_{w}=T_{H}\left(\frac{1+\sqrt{1-\gamma}\cosh\left(2\lambda y_{w}
\right)}{\sqrt{1-\gamma}\sinh\left(2 \lambda
y_{w}\right)}\right)\label{ttolman}\hspace{1ex}.\ee The entropy of
the two-dimensional charged stringy black hole is given by \be
S=-\left(\frac{\partial F}{\partial y_{w}}\right)_{\lambda, D,
Q_{el}}\left(\frac{\partial T_{w}}{\partial
y_{w}}\right)^{-1}_{\lambda, D, Q_{el}}\ee where the free energy
$F=F(\lambda, D, T_{w}, y_{w})$ is \be F=-4\lambda D \coth
\left(2\lambda y_{w}\right)\ee and thus the explicit expression
for entropy is \be S=4 \pi
e^{\phi_{w}}\frac{\left(1+\frac{1}{\sqrt{1-\gamma}}\right)}{\left(1+\frac{\cosh\left(2\lambda
y_{w} \right)}{\sqrt{1-\gamma}}\right)}\label{entropy2}\ee where
equations (\ref{tg}), (\ref{dilaton2}), (\ref{dcharge}) and
(\ref{ttolman}) have been substituted in (\ref{entropy2}). It is
clear that even if someone takes into consideration the fact that
the entropy is affected by a change in the boundary value
$\phi_{w}$ of the dilaton field  and thus affected by the
variation of $\alpha$ as seen from equation (\ref{alpha2}), the
entropy is getting smaller with respect to an increase in
$\gamma$, i.e. an increase in the electric charge $e$. Therefore,
although we have concerned the two-dimensional charged stringy
black hole in a thermal environment, i.e. in the ``box'', an
increase in the electric charge causes the entropy of the above
mentioned gravitational background to be smaller. The violation of
the second law of thermodynamics seems to be inevitable.
\par
In summary, we have viewed a two-dimensional stringy black hole as
an isolated object and also as contained in a thermal environment.
In both cases, it was shown that an increase of the electric
charge $e$ leads to an expected decrease of the temperature and to
an unexpected decrease of the entropy of the two-dimensional
charged stringy black hole. The latter result implies that varying
$e$ theories run the risk of violating the second law of
thermodynamics. \par\noindent Therefore, we believe that our
results are a signal of the specific model, c.f.
\cite{flam,carlip2,carlip1,fair}, meaning that results derived in
the context of black hole thermodynamics will always be
model-dependent and will not lead to any constrains on varying
constant theories.  Of course, one could claim that the results
derived here are  a support for the arguments of Davies et al
\cite{davies} that black holes are able to discriminate between
the varying $e$ and varying $c$ theories. But in this case, one
has a priori to believe that experiments measure dimensional
quantities which is completely erroneous \cite{duff1}.
\par\noindent Finally, a couple of points are in order. Firstly,
the analysis presented here has taken for granted the increase of
the fine structure $\alpha$ in time although a confirmation of
this variation is still lacking in astrophysical observations.
Secondly, we have assumed that the variation of the fine structure
$\alpha$ is due to an increase in the electric charge $e$
\footnote{It should be pointed out once more that by variation of
the electric charge $e$ we mean the variation of the dimensionless
quantities that depend on $e$. The same analogously holds for the
other dimensional constants $c$ and $\hbar$. }. There is also the
possibility that the variation of $\alpha$ could be due to a
variation in the speed of light $c$ or due to a variation of the
Planck's constant $\hbar$ or even due to a simultaneous variation
of some of the aforesaid constants \cite{barrow5}.
\section*{Acknowledgements}
The author is grateful to Professors  R. Emparan, D. Espriu and
J.G. Russo for enlightening comments and useful conversations. The
author is also indebted to Dr. A. Ghosh for useful correspondence.
Last but not least, special thanks are due to the referee for
constructive criticism.  This work has been supported by the
European Research and Training Network ``EUROGRID-Discrete Random
Geometries: from Solid State Physics to Quantum Gravity"
(HPRN-CT-1999-00161).

\end{document}